\documentclass[aps,prl,twocolumn,showpacs,superscriptaddress,groupedaddress]{revtex4}
\usepackage{graphicx}  
\usepackage{dcolumn}   
\usepackage{bm}        
\usepackage{amssymb}   

\usepackage{bbm}
\usepackage{amsmath, amsthm, amsfonts, amssymb}
\usepackage{mathrsfs} 
\usepackage{minibox}
\usepackage{graphics}
\usepackage{epsfig}
\usepackage{xcolor}
\usepackage{array}
\usepackage{tikz}
\usepackage{tikz-cd}
\usetikzlibrary{arrows,calc}

\def\b{\beta}

\def\d{\delta}
\def\ve{\varepsilon}
\def\m{\mu}

\def\l{\lambda}
\def\L{\Lambda}

\def\wt{\widetilde}

\newcommand{\be}{\begin{equation}}
\newcommand{\ee}{\end{equation}}
\newcommand{\bea}{\begin{eqnarray}}
\newcommand{\eea}{\end{eqnarray}}
\newcommand{\bal}{\begin{aligned}}
\newcommand{\eal}{\end{aligned}}
\newcommand{\eq}[1]{Eq.~(\ref{#1})}

\begin{document}

\title{Symmetry breaking and RG flows with higher dimensional operators}
\author{Nikos Irges, Fotis Koutroulis}
\affiliation{Department of Physics, National Technical University of Athens, GR-15780 Athens, Greece}

\date{\today}

\begin{abstract} We discuss the role of higher dimensional operators in the spontaneous breaking of internal symmetry and scale invariance,
in the context of the Lorentz invariant scalar field theory. Using the $\ve$-expansion we determine phase diagrams and demonstrate that 
(un)stable RG flows computed with a certain basis of dimension 6 operators in the Lagrangian map to (un)stable RG flows of another basis related to the first by field redefinitions.
Crucial is the presence of reparametrization ghosts if Ostrogradsky ghosts appear.

\end{abstract}

\maketitle

\tikzset{
    photon/.style={decorate, decoration={snake}, draw=black},
    electron/.style={draw=black, postaction={decorate},
        decoration={markings,mark=at position .55 with {\arrow[draw=black]{>}}}},
    gluon/.style={decorate, draw=black,
        decoration={coil,amplitude=4pt, segment length=5pt}} 
}

\section{Introduction}\label{Intro}
 
The Higgs potential responsible for the spontaneous breaking of gauge symmetry breaks also scale invariance via an explicit mass term, already at the classical level.
Sometimes the breaking of internal symmetries is correlated with the breaking of scale invariance at the quantum level. Such is the case of the Coleman-Weinberg model  \cite{Coleman} where
in the absence of a relevant (mass) operator in the Lagrangian quantum effects induce the simultaneous breaking of gauge and scale symmetries.

A simplified version of spontaneous symmetry breaking (SSB) can be studied in the context of a scalar field theory with a kinetic term and a potential 
containing both mass and quartic self-interaction terms. 
The internal symmetry in this case is the global $\mathbb{Z}_2$ transformation $\phi\to -\phi$.
When the signs of the two terms in the potential are opposite and the overall sign of the potential is right, SSB is triggered already at the classical level by
the scalar field acquiring a nonzero vacuum expectation value (vev). 
Without the mass term this model does not have a phase where the $\mathbb{Z}_2$ symmetry is broken neither in the classical limit nor at the quantum 
level, as far as perturbation theory is concerned. Scale invariance on the other hand generically does break even in the massless limit by the coupling developing non-zero $\b$-function.

Another generic consequence of quantization is the appearance of classicaly irrelevant, higher dimensional operators (HDOs) in the action, suppressed by appropriate powers of
a dimensionful scale $\L$.  
To the extent that they are quantum in nature, the breaking of scale invariance induced by such operators is spontaneous and not explicit.
In a different regularization scheme we could suppress them by the regulating scale itself but this should not matter.
Specific properties of the phase diagram may be regularization dependent but global properties are expected to be regularization independent. 
For example, the topology of the phase diagram and global properties of the Renormalization Group (RG) flow lines in it are expected to be such. 
Meanwhile the dimensionless couplings that multiply HDOs run with the regulating scale in a way dictated by the regularization scheme, as any other
coupling. The running of all couplings, associated with relevant, marginal or irrelevant operators is accordingly correlated, resulting in a multidimensional phase diagram.
An interesting scenario is one where for some reason the Lagrangian contains HDOs but
does not contain mass terms or classicaly marginal operators. Then, if the phase diagram
develops a $\mathbb{Z}_2$ broken phase at the quantum level, perturbation theory should be able to detect it via the presence of these HDOs. 
There is a particularly interesting interpretation of this picture, revealed by the observation that field redefinitions leave the S-matrix invariant.
This allows to rotate in field space onto a basis with no higher derivative terms in the Lagrangian, which makes the $\mathbb{Z}_2$  breaking
phase more transparent.

In order to access these phenomena we focus on a scalar field theory with HDOs of dimension 6 and 8. We use Dimensional Regularization with regulating scale $\m$ and the expansion parameter $\ve$,
where $d=4-\ve$ is the space-time dimension. The $\ve\to 0$ limit defines the four-dimensional theory. 
We will also examine the cases $\ve=\pm 1$ which correspond to the three-dimensional and five-dimensional versions of the model. The reason is that in these cases the
phase diagram may contain Wilson-Fisher (WF) fixed points where scale invariance is restored, to the order that perturbation theory is used.
We restrict our analysis to one-loop.
Detailed physical properties could still be influenced by the subtraction scheme.
Here, since we restrict our attention to the RG flows and the phase diagram that are independent of the finite parts of the loop diagrams,
we will absorb them into the counter-terms without loosing generality.

\section{Review of the ${\cal L}^{(4)}$ model}\label{2-4}

We first look at the standard case, reconstructing known results reviewed in \cite{Peskin}. 
In the following, a subscript 0 on a Lagrangian stands for "bare", in which case all fields and couplings contained in it inherit the subscript.
At 1-loop this is trivial for the field $\phi$ since it has a vanishing anomalous dimension. For all other fields and couplings the renormalization process
starts by writing, for example for a coupling $c$: $c_0=c+\d c$ with $c$ the renormalized coupling and $\d c$ its counter-term.

The information relevant to the Lagrangian 
\be\label{L4}
{\cal L}^{(4)} = -\frac{1}{2} \phi \Box \phi  -\frac{1}{2}  \phi\, m^2 \phi  -\frac{\l}{4!}\phi^4
\ee
which now includes the renormalized fields and couplings, is encoded in the one-loop $\b$-functions
\be\label{bc1m}
\b_{m^2}(m^2,\l) = -2 m^2 + \frac{\l m^2}{16 \pi^2}, \hskip .1 cm \b_{\l}(\l) = -\ve \l + \frac{3\l^2}{16 \pi^2}
\ee
It is understood that one starts from ${\cal L}_0^{(4)}$, then shifts bare parameters as stated and absorbs the divergences in the counter-terms.
The counter-terms then determine the above $\b$-functions and one is left with the renormalized Lagrangian in \eq{L4}.
The phase diagram on the $m^2$ vs $\l$ plane
has two types of fixed points. One is the Gaussian fixed point defined by $m^2_{\bullet}=\l_{\bullet}=0$ and is denoted as $\bullet$.
The other is the Wilson-Fisher fixed point determined by the vanishing of both $\b$-functions with at least one non-zero coupling, denoted as $\star$. 
Here this happens for $m^2_{\star}=0$ and $\l_{\star}=\frac{16\pi^2}{3}\ve$ and it is present when $\ve \ne 0$. 
Restricting to the regime where a stable $\mathbb{Z}_2$ breaking phase is observed (with our sign conventions when $\l > 0$ and $m^2 < 0$),
in $d=4$ and $d=5$ there is only one class of RG flow lines with a Landau pole $\m_L$ in the UV, where $\l$ diverges. 
The mass which is a relevant operator grows in the IR where $\l$ is small and this is one way to expose its unnaturalness.
In $d=3$ there is a flow without Landau pole emanating from the Gaussian point where the mass and $\l$ vanish, towards the WF point in the IR where the mass diverges.
This is the continuum branch of the broken phase. Extending the validity of the $\ve$-expansion beyond the WF point
yields RG flow lines with a Landau pole in the UV, where $\l$ diverges, and that approach the WF point towards the IR with a diverging mass.
This is the Landau branch and it has a similar naturalness issue with the $d=4,5$ cases. 
In all but one cases the mass operator when non-zero breaks scale invariance, as do the $\b$-functions away from the fixed points.
The one case where the $\mathbb{Z}_2$ is broken with scale invariance restored while the system is interacting, is the $d=3$ WF point.
The physical scalar mass in the stable, broken phase is $m_{\phi}^2 = \frac{1}{3} \l v^2$.

\section{Inserting dimension-6 operators}\label{2-4-6}

Let us consider the Lagrangian ${\cal L}_0 = {\cal L}_0^{(4)}+{\cal L}_0^{(6)}$ where
\bea\label{L46}
{\cal L}_0^{(6)} &=& \frac{c_{1,0}^{(6)}}{ 4! \L^2} \phi^2 \Box \phi^2 + \frac{c_{2,0}^{(6)}}{ 2 \L^2} \phi \Box^2 \phi + \frac{c_{3,0}^{(6)}}{ 6! \L^2}  \phi^6 
\eea
and ask about its symmetry breaking properties. ${\cal L}_0$ features the three dim-6 operators $c^{(6)}_{i,0} O^{(6)}_{i,0}, i=1,2,3$
in the order they appear in ${\cal L}_0^{(6)}$ but only one of them is independent. 
It is well known for example that by using the classical equations of motion one can eliminate any two of these operators.
Alternatively, one can perform appropriate field redefinitions to arrive at the same result and this is the method we will use here,
by making sure in addition that the reduction of the operator basis is consistent at the 1-loop level.
Moreover, $O^{(6)}_{1,0}$ can result
in superluminal propagation when $c^{(6)}_{1,0}<0$ \cite{Nicolis} and $O^{(6)}_{2,0}$ contains an Ostrogradsky ghost (the O-ghost) \cite{Richard}
as it adds an extra, ghost-like pole to the propagator. Let us call this set of operators the G-basis, since it is the most general basis,
consistent with the Lorentz and the internal symmetry, up to total derivatives. Due to these issues, the analysis is more clear in the 
basis where only $O^{(6)}_{3,0}$ appears, the W-basis, by analogy to the Warsaw basis of dim-6 operators in the Standard Model \cite{Warsaw}
which is ghost free. Thus our first task is to establish a connection between the G and W-bases.

In order to make sense of Eq. (\ref{L46}) we must add a sector
whose purpose is to cancel the pole of the O-ghost:
\be\label{Reg}
{\cal L}_{Rg,0}=-\frac{1}{2}{\bar \chi}_0 \Box \chi_0 + \frac{1}{2}m_\chi^2 {\bar \chi}_0 \chi_0 -\frac{1}{4}\l_{\chi,0} {\bar \chi}_0 \chi_0 \phi^2
\ee
We will call the Grassmann field $\chi$ the R-ghost for reasons that will become clear. 
We denote the 1-loop correction to the 2, 4 and 6-point function of $\phi$, ${\cal M}_{2,\phi}$, ${\cal B}_{4,\phi}$ and ${\cal B}_{6,\phi}$
respectively, the 2-point function of $\chi$, ${\cal M}_{2,\chi}$ and the correction to the $\phi$-$\chi$ vertex ${\cal B}_{4,\chi}$.
We also introduce the counter-terms $\delta \phi$, $\delta \chi$, $\delta m$, $\delta m_\chi $, $\d\l$, $\d c^{(6)}_i$ and $\d \l_\chi$, 
after which bare quantities are replaced by renormalized ones.

The conditions that the pole in the $\chi$-propagator cancels the unwanted pole of the $\phi$ propagator are
$m^2=0$ and $m_\chi^2=\L^2 /c^{(6)}_2$ at the classical level and $\d\phi=0$ and $\d\chi=-{\cal M}_{2,\chi}/(p^2+m_\chi^2)$ at the 1-loop level.
We also have the identity $\d m_\chi/m_\chi^2=-\d c_2/c_2$.
The condition $m^2=0$ forces $\d m = {\cal M}_{2,\phi}$.
The other conditions that render ${\cal L}$ renormalized are
\be
\d \left(\l + \frac{p^2}{\Lambda^2}c^{(6)}_1 \right) = {\cal B}_{4,\phi}, \,\,\, \d c^{(6)}_3 = - \L^2{\cal B}_{6,\phi}\nonumber
\ee
and $\d\l_\chi = {\cal B}_{4,\chi}-\l_\chi \d\chi$. 

Substituting the values of the Feynman diagrams (see Appendix) we obtain the counterterms (in units of $1/(16\pi^2 \ve)$)
\bea\label{deltauntilded}
&& \d\l = 3\l^2 + 3\frac{c_3^{(6)}m_\chi^2}{\L^2}, \hskip .75cm \d c_3^{(6)} = - 5\l c_3^{(6)} \nonumber\\
&& \frac{p^2}{\L^2} \d c_1^{(6)}= 
- 6 ( \l^2 + \frac{ \l_\chi^2}{4} 
+ c_{3}^{(6)}  \frac{ m_{\chi}^2 }{\L^2} )\nonumber\\
&& \d m = (\l + \l_\chi) m_\chi^2 + \frac{c_1^{(6)}}{c_2^{(6)}} p^2
\eea
An observation is that the coupling $\l_\chi$ does not play any physical role and 
can be set to zero without loss of generality. Notice that the renormalized ghost-$\phi$ coupling is defined at a specific subtraction scale $\m_{\rm R}$. Then holds that $\l_\chi \equiv \l_\chi(\m_{\rm R})$
but also we have $\l_\chi(\m) \sim \l_\chi(\m_{\rm R})$ 
which shows that if the renormalized coupling is zero then the R-ghosts are decoupled independently of the subtraction scale. In fact this decouples the R-ghost from $\phi$.
With the above conditions the renormalized Lagrangian becomes
\bea\label{Lagmult}
{\cal L}&=&-\frac{1}{2}\Bigl[-(p^2+m_\chi^2)(\frac{p^2}{m_\chi^2}\phi^2+{\bar \chi}\chi)\Bigr] - \frac{\l +\frac{p^2}{\Lambda ^2} c^{(6)}_1 }{4!}\phi^4 \nonumber\\
&+& \frac{c^{(6)}_3}{6!\L^2}\phi^6 
+\frac{\d c^{(6)}_2}{ 2 c^{(6)}_2}\left(\frac{p^4}{m_\chi^2}\phi^2-m_{\chi}^2{\bar \chi}\chi\right)
\eea
The pole cancellation condition forces $\d c^{(6)}_2$  to remain in the Lagrangean so we can attempt to use it as a Lagrange multiplier,
in which case the fluctuations of $g(x) \equiv \frac{p^4}{m_\chi^2} \phi(x)^2 - m_\chi^2 {\bar \chi}\chi(x)$ are constrained to vanish.
Then it can be integrated out, giving ${\bar \chi}\chi=p^4/m_\chi^4 \phi^2$, which
determines a composite field $\Phi$ as
\be\label{Phi}
\Phi^2 = \frac{p^2}{m_\chi^2}(1+ \frac{p^2}{m_\chi^2})\phi^2
\ee
in terms of which ${\cal L}$ becomes
\be\label{LPhi}
{\cal L}=-\frac{1}{2}\Bigl[\Phi(\Box-{m^2_\chi})\Phi\Bigr]-\frac{\lambda'}{4!}\Phi^4+\frac{c^{(6)'}_{3}}{6!\Lambda^2}\Phi^6
\ee
with $\l'$ and $c^{(6)'}_{3}$ easily obtained using Eq. (\ref{Phi}). Since Eq. (\ref{LPhi}) hides a non-local form it is not particularly useful
for computations. 
We keep in mind its form though which suggests that the operators $O^{(6)}_1$ and $O^{(6)}_2$ are indeed redundant.
Notice finally that the pole has shifted to a mass $-m_\chi^2$ for the non-local field $\Phi$, while the non-zero $\d m$
determines a non-trivial counter-term in a similar fashion as in the Coleman-Weinberg model \cite{Coleman}.

There is a way to bring ${\cal L}$ to a local form. Start from 
\be\label{Ltilded}
{\wt {\cal L}}=-\frac{1}{2}\Bigl[\phi(\Box+{{\wt m}}^{2})\phi\Bigr]-\frac{\wt\lambda}{4!}\phi^4+\frac{{\wt c}^{(6)}_{3}}{6!\Lambda^2}\phi^6
\ee
and perform the field redefinition
\be\label{phitr}
\phi\to \phi + \frac{x}{\L^2} \Box \phi + \frac{y}{\L^2} \phi^3 
\ee
We then obtain the Lagrangean ${\cal L}^{(4)}+{\cal L}^{(6)}$ up to $O(1/\L^2)$ plus, taking into account the Jacobean of the transformation, 
a ghost action, equivalent to ${\cal L}_{R}g$ of Eq. (\ref{Reg}).
The freedom inherent in this operation is fixed by choosing a diagonal gauge, where $(\partial^{2n}\phi)'$ is considered to be function of only $\partial^{2n}\phi$.
Therefore, $\chi$ becomes the reparametrization ghost associated with Eq. (\ref{phitr}), justifying its name.
The coefficients are fixed to $x=-1/2 c^{(6)}_2$, $y=-1/4!(c^{(6)}_1-2{\wt \l}c^{(6)}_2)$ and
the couplings in the un-tilded and tilded bases are related via
\bea
m^2&=& {\wt m}^2 (1+\frac{c_2^{(6)}{\wt m}^2}{\L^2}), \,\, \l = {\wt \l}-\frac{{\wt m}^2}{\L^2} (c_1^{(6)}-4c_2^{(6)}{\wt \l})\nonumber\\
c_3^{(6)}&=& {\wt c}_3^{(6)} + 5{\wt \l} (c_1^{(6)}-2c_2^{(6)}{\wt \l})\nonumber\\
\l_\chi &=& \frac{6y}{x} (1+\frac{c_2^{(6)}{\wt m}^2}{\L^2})\nonumber
\eea
In order that the reparametrization ghost decouples, we need $\l_\chi=0$ which can be achieved when $y=0$ or ${\wt m}^2=-m_\chi^2$.
This is an expected fact since $\l_\chi$ is connected with the ultra-local part of Eq. (10) which has a trivial role.
The former happens if $c^{(6)}_1=2{\wt \l}c^{(6)}_2$ and the latter ensures that $m^2=0$ and
we need both in order to get an equivalent to Eq. (\ref{LPhi}) theory.
With these constraints, we have $\l= - {\wt\l}$ and $c_3^{(6)}= {\wt c}_3^{(6)}$
and the $\b$-functions in the two bases agree.
Note that the cancellation of the dangerous pole does not happen via a diagrammatic cancellation as it is the case with Faddeev-Popov ghosts for example.
It is rather imposed on the system by requiring that the pole of the $\phi$ propagator be the same with the pole of $\chi$, which by itself is a necessary
but not sufficient condition. A direct relation between the fields themselves is required in order for it to be also sufficient. This condition is automatically
generated by the way that $\d c^{(6)}_2$ appears in \eq{Lagmult} as a Lagrange multiplier. 

To close the cycle we renormalize Eq. (\ref{Ltilded}) directly. The diagrams in the W-basis are the same with those of the G-basis but with the ones that involve $\chi$ not contributing.
The result (in units of $1/16\pi^2\ve$) for the counter-terms is $\d {\wt m}=\wt \l {\wt m}^2$,
$\d \wt\l=3{\wt \l}^2 - 3 \frac{\wt c_3^{(6)} \wt m^2}{\L^2}$ and $\d{\wt c_3^{(6)}} = 5 \wt \l \wt c_3^{(6)}$, in agreement with Eq. (\ref{deltauntilded}).
The $\b$-functions of the W-basis are then $\b_{\wt m^2}=-2{\wt m}^2+\wt \l {\wt m^2}/16\pi^2$ and
\be\label{bwtlbwtc6}
\b_{\wt \l} = -\ve {\wt \l} + \frac{3{\wt \l}^2}{16\pi^2} - \frac{3{\wt c}_3^{(6)}{\wt m}^2}{16\pi^2\L^2} , \,\, \b_{ \wt c_3^{(6)} } = -2\ve \wt c_3^{(6)} + \frac{5 \wt \l \wt c_3^{(6)}}{16 \pi^2}  
\ee 
We now set ${\wt m}^2=0$ and analyze the phase diagram.
In $d=4$ the ${\wt \l}=0$ axis is a line of scale invariant theories labelled by the value of ${ \wt c_3^{(6)} }$.
Upon absorbing all finite parts of diagrams in the counter-terms, the renormalized potential assumes the form
\be
\wt V^{(6)} = \frac{\wt \l}{4!} \phi^4 - \frac{ \wt c_3^{(6)}}{6!\L^2}  \phi^6
\ee
The $\mathbb{Z}_2$ breaking is stable when ${\wt \l}<0$ and ${ \wt c_3^{(6)}}<0$. 
The physical scalar mass in this phase is
$m_{\phi}^2 =  \frac{1}{3} |{\wt \l}| v^2$ with  ${v}^2 = 20 \frac{\wt \l}{\wt c_3^{(6)}} \L^2$.
Along a $v^2={\rm const.}$ line the mass induced by SSB sees the Landau pole of $\wt \lambda$ and diverges in the UV.

\begin{figure}[!htbp]
\centering
\includegraphics[width=7cm]{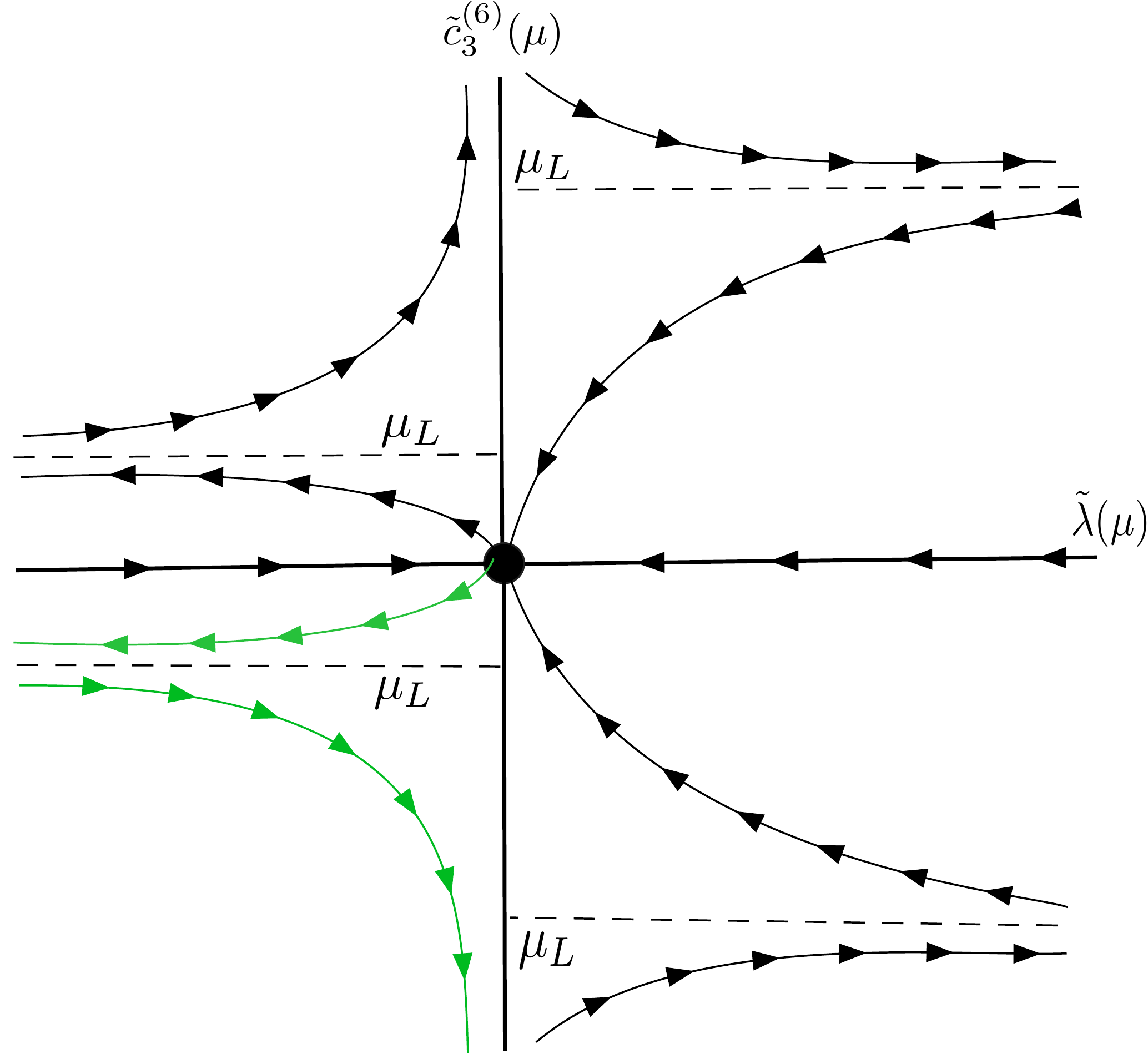}
\caption{\small RG flows for ${\wt {\cal L}}$ in Eq. (\ref{Ltilded}), in $d=4$. The ${\wt \l}, {\wt c_3^{(6)}} < 0$ (green) flow corresponds to the broken phase.
\label{d4c63lpo}}
\end{figure}

Scale invariance is broken by quantum fluctuations at a generic point in the phase diagram and it is restored only at the UV $\bullet$ fixed point.
Comparing Eqs. (\ref{bc1m}) and (\ref{bwtlbwtc6}) one sees that 
${ \wt c_3^{(6)}}$ acts as a mass operator in $d=3$ and as an inverse mass operator in $d=5$.
In a bit more detail, in their respective broken phases, the $d=4$ flows of ${\wt {\cal L}}$ are the same with those of the ${\cal L}^{(4)}$ model for ${\l}<0$.
The $d=3$ flows of ${\wt {\cal L}}$ are the same as the 
$d=3$ flows of ${\cal L}^{(4)}$. The $d=5$ flow of ${\wt {\cal L}}$ on the other hand starts off $\bullet$ in the IR where
both couplings vanish and tends to a constant $\wt\l$ and a diverging ${ \wt c_3^{(6)}}$ in the continuum limit.
The flows for ${\wt {\cal L}}$ in $d=4$ are depicted in Fig. \ref{d4c63lpo}, 
with the arrows pointing to the IR.
Note that $\L$ is not an independent scale in the unbroken phase either. In the UV and IR limits, if it is not related to a Landau pole
then it can be removed (at the WF point). Anywhere in between a relation between $\L$ and $\m$ can be established order by order
in perturbation theory by a systematic but painful process.

%

\section{Dimension-8 operators}\label{6-8}

Next we add to ${\cal L}_0^{(4)}+{\cal L}_0^{(6)}$ its most general (again up to total derivatives) dim-8 extension
\be
{\cal L}_0^{(8)} = \frac{c_{1,0}^{(8)}}{ 6! \L^4} \phi^3 \Box \phi^3 + \frac{c_{2,0}^{(8)}}{ 4! \L^4} \phi^2 \Box^2 \phi^2 + 
\frac{c_{3,0}^{(8)}}{2 \L^4} \phi \Box^3 \phi + \frac{c_{4,0}^{(8)}}{ 8! \L^4} \phi^8\nonumber
\ee
and rotate to a basis where only the $\phi^6$ and $\phi^8$ from the HDOs are kept.
In the Appendix we show the rotation needed to arrive at this form (for simplicity for the special case $m^2=\l=0$).
We will not go through the consistency checks of the previous section which are now more complicated.
Here we only ask if we can have a quantum breaking of internal and scale symmetry without 
a mass term or a quartic interaction.
To lighten the notation we drop the tildes specifying the rotated basis. 

Renormalization determines
\be\label{bfmlc6c8}
\vec{\b}  = \begin{bmatrix}
  -2 m^2 + \frac{\l m^2}{16 \pi^2}      \\
 -\ve \l + \frac{3  \l^2}{16 \pi^2}  - \frac{3  c_3^{(6)}  m^2}{16 \pi^2\L^2}  \\
 -2\ve  c_3^{(6)} + \frac{5  \l  c_3^{(6)} }{16 \pi^2} + \frac{5  c_4^{(8)}}{16 \pi^2} \frac{ m^2 }{\L^2}   \\
 -3\ve  c_4^{(8)} + \frac{7 \l  c_4^{(8)} }{16 \pi^2} + \frac{7 ( c_3^{(6)})^2 }{16 \pi^2}  \\
\end{bmatrix} 
\ee
and ${\cal L}^{(8)} = -\frac{1}{2} \phi \Box \phi  - {V}^{(8)} $ with
\be\label{wtV1lc6c8}
{V}^{(8)} =  \frac{1}{2}  m^2 \phi^2 + \frac{ \l}{4!}\phi^4 - \frac{ c_3^{(6)}}{6! \L^2} \phi^6 - \frac{ c^{(8)}_4}{8! \L^4} \phi^8 \nonumber
\ee
Let us now fix ${ m}^2={\l}=0$. 
The potential is stable when ${c}_4^{(8)}<0$ and when in addition ${ c}_3^{(6)}>0$ a vev
breaks the $\mathbb{Z}_2$ and scale invariance, both restored only in the continuum limit. The resulting scalar mass is 
$m_\phi^2 = \frac{7}{10} \frac{(c_3^{(6)})^2 }{|c_4^{(8)}|} v^2$ with ${ v}^2 = {42} \frac{ c_3^{(6)}}{|c_4^{(8)}|} \L^2$.

In $d=4$, when $c_3^{(6)} = 0$ neither coupling runs with $\m$ and the phase diagram has a WF-line 
with points labelled by $c_4^{(8)}$ which ends on a Gaussian fixed point where ${c_{3\bullet}^{(6)}} = { c_{4\bullet}^{(8)}}=0$. 
On the other hand, if $c_3^{(6)} \ne 0$ then $c_4^{(8)} $ has a non-trivial RG flow at a constant distance equal to $c_3^{(6)}$ from the WF-line
that forces it to vanish in the IR at some scale $\m_L$ and to diverge in the UV.
The stable, broken branch of the flow (when $c_4^{(8)} < 0$) is for $0<\m<\m_L$ with $\m_L=\m_Re^{16\pi^2 \frac{v^2}{10m_\phi^2}}$ and $\m_R<\m_L$
the renormalization scale. This imposes on $m_\phi^2$ a UV Landau pole. The difference here is that the branches below and above the Landau Pole are
continuously connected, rendering SSB overall unstable.
In $d=3$ the Gaussian is a UV fixed point while the couplings both diverge in the IR. In $d=5$ the flows are the reverse.

\section{Conclusion}

We have considered a simple case in the context of Generalized Effective Field Theories, which are effective field theories 
where the Wilson coefficients develop their own, independent scale dependence.
We analyzed the one-loop relation between RG flows related by field redefinitions in the scalar field theory,
up to operator dimension 6. Our results could be relevant to computations performed in the Standard Model, particularly
in bases related to the Warsaw basis by the use of equations of motion.
In particular we showed how a basis that contains an Ostrogradsky ghost can be consistently defined with the addition of 
an extra sector of ghost-like fields. These ghosts are then interpreted as reparametrization fields associated with the non-trivial Jacobean 
originating from the field redefinition that is necessary to eliminate the dangerous operator.
We showed that the construction does not work at the classical level, it requires quantum effects and  renormalization. 
We showed how internal and scale symmetry break spontaneously in the pure polynomial basis without a mass term at operator dimension 6 and without mass and
quartic interactions at dimension 8. We also determined the RG flows of the operator dimension 8 model in one basis of operators.
This could be relevant to quantum effective actions that do not generate explicit mass or marginal potential terms.
We intend to present the details of the calculations involved in this letter in \cite{IrgesFotis4}.
Previous work on related issues can be found in \cite{phi6quantum}, each of which may have some overlap with our analysis.

\section{Acknowledgement}
We would like to thank A. Kehagias for discussions.

\section{Appendix}
\begin{appendix}

%
The 1-loop quantities needed for the renormalization of ${\cal L}$ are:
\bea
{\cal M}_{2,\phi} &=& -  \frac{1}{2} \left(  \l +  c_{1}^{(6)} \frac{q^2} {\L^2}  \right) A_0(-m^2_\chi) - \frac{ \l_{\chi}}{2}  \, A_0( - m_{\chi}^2 ) \nonumber\\
{\cal M}_{2,\chi} &=& - \frac{ \l_{\chi}}{2}  \, A_0( - m_{\chi}^2 ) \nonumber\\
{\cal B}_{4,\phi} &=& \sum_{q^2=s,t,u} \frac{(\l+\frac{q^2}{m_\chi^2}\frac{c_1^{(6)}}{c_2^{(6)}})^2}{2} F(q^2,m_\chi)  \nonumber\\
&-& \sum_{q^2=s,t,u} \frac{ \l_{\chi}^2}{4}  B_0(q^2,m_\chi,m_\chi) + \frac{3}{2} \frac{c_3^{(6)}}{m_\chi^2 c_2^{(6)}}A_0(-m^2_\chi) \nonumber\\
\L^2 {\cal B}_{6,\phi} &=& - \sum_{q^2=s_1,\cdots,s_5} \frac{(\l+\frac{q^2}{m_\chi^2}\frac{c_1^{(6)}}{c_2^{(6)}}) c_3^{(6)} }{2}  F(q^2,m_\chi)\nonumber\\
{\cal B}_{4,\chi} &=& \sum_{q^2 = s,t}  \frac{\l_{\chi}}{4}  \left(  \l + c_{1}^{(6)} \frac{q^2} {\L^2}  \right)  F(q^2, m_{\chi}^2) \nonumber
\eea
where 
\bea
&&A_0(m^2)=m^2[2/(16\pi^2\ve ) +\cdots ]\nonumber\\
&& B_0(p^2,m_1,m_2)=2/(16\pi^2\ve ) +\cdots\,\, {\rm when}\,\,  m_{1,2}\ne 0  \nonumber
\eea
with the dots representing finite terms and
\bea
F(q^2,m_\chi) &=& B_0(q^2,0,0) - B_0(q^2,0,m_\chi) - B_0(q^2,m_\chi,0) \nonumber\\
&+& B_0(q^2,m_\chi,m_\chi) \nonumber
\eea
with $B_0(q^2,0,0)\equiv 0$. 

The redefinition on $\phi$ 
\be
\phi\to \phi + \frac{x}{\L^2} \Box \phi + \frac{y}{\L^2} \phi^3 + \frac{z}{\L^4} \Box^2 \phi + \frac{u}{\L^4} \phi^2 \Box \phi + \frac{w}{\L^4} \phi^5\nonumber
\ee
in the $m^2=\l=0$ limit, eliminates all higher derivative terms from the dim-8 Lagrangian for
\bea
x &=& c^{(6)}_2,\hskip .5cm y = c^{(6)}_1,\hskip .5cm z= 5 c^{(6)}_1 c^{(6)}_2 + c^{(8)}_2,  \nonumber\\
u&=&\frac{3}{2} (c^{(6)}_2)^2 + c^{(8)}_3,\hskip .5cm  w= \frac{7}{2} (c^{(6)}_1)^2 + 6 c^{(6)}_2 c^{(6)}_3 + c^{(8)}_1 \nonumber
\eea
provided that total derivative terms are dropped.
After a field redefinition, if necessary, the kinetic term is brought to canonical form by an extra redefinition.

\end{appendix}


\end{document}